\documentstyle[twoside,fleqn,espcrc2]{article}
\newcommand{\beq}{\begin{equation}}
\newcommand{\eeq}{\end{equation}}
\newcommand{\bea}{\begin{eqnarray}}
\newcommand{\eea}{\end{eqnarray}}
\newcommand{\bda}{\begin{eqnarray*}}
\newcommand{\eda}{\end{eqnarray*}}

\newcommand{\dg}{\dagger}

\newcommand{\Id}{1\!\!1}
\newcommand{\Tr}{{\rm Tr}}
\newcommand{\Epsilon}{{\cal E}}
\newread\epsffilein    
\newif\ifepsffileok    
\newif\ifepsfbbfound   
\newif\ifepsfverbose   
\newdimen\epsfxsize    
\newdimen\epsfysize    
\newdimen\epsftsize    
\newdimen\epsfrsize    
\newdimen\epsftmp      
\newdimen\pspoints     
\def\epsfbox#1{\global\def\epsfllx{72}\global\def\epsflly{72}%
   \global\def\epsfurx{540}\global\def\epsfury{720}%
   \def\lbracket{[}\def\testit{#1}\ifx\testit\lbracket
   \let\next=\epsfgetlitbb\else\let\next=\epsfnormal\fi\next{#1}}%
\def\epsfgetlitbb#1#2 #3 #4 #5]#6{\epsfgrab #2 #3 #4 #5 .\\%
   \epsfsetgraph{#6}}%
\def\epsfnormal#1{\epsfgetbb{#1}\epsfsetgraph{#1}}%
\def\epsfgetbb#1{%
\openin\epsffilein=#1
\ifeof\epsffilein\errmessage{I couldn't open #1, will ignore it}\else
   {\epsffileoktrue \chardef\other=12
    \def\do##1{\catcode`##1=\other}\dospecials \catcode`\ =10
    \loop
       \read\epsffilein to \epsffileline
       \ifeof\epsffilein\epsffileokfalse\else
          \expandafter\epsfaux\epsffileline:. \\%
       \fi
   \ifepsffileok\repeat
   \ifepsfbbfound\else
    \ifepsfverbose\message{No bounding box comment in #1; using defaults}\fi\fi
   }\closein\epsffilein\fi}%
\def\epsfclipstring{}
\def\epsfsetgraph#1{%
   \epsfrsize=\epsfury\pspoints
   \advance\epsfrsize by-\epsflly\pspoints
   \epsftsize=\epsfurx\pspoints
   \advance\epsftsize by-\epsfllx\pspoints
   \epsfxsize\epsfsize\epsftsize\epsfrsize
   \ifnum\epsfxsize=0 \ifnum\epsfysize=0
      \epsfxsize=\epsftsize \epsfysize=\epsfrsize
      \epsfrsize=0pt
     \else\epsftmp=\epsftsize \divide\epsftmp\epsfrsize
       \epsfxsize=\epsfysize \multiply\epsfxsize\epsftmp
       \multiply\epsftmp\epsfrsize \advance\epsftsize-\epsftmp
       \epsftmp=\epsfysize
       \loop \advance\epsftsize\epsftsize \divide\epsftmp 2
       \ifnum\epsftmp>0
          \ifnum\epsftsize<\epsfrsize\else
             \advance\epsftsize-\epsfrsize \advance\epsfxsize\epsftmp \fi
       \repeat
       \epsfrsize=0pt
     \fi
   \else \ifnum\epsfysize=0
     \epsftmp=\epsfrsize \divide\epsftmp\epsftsize
     \epsfysize=\epsfxsize \multiply\epsfysize\epsftmp   
     \multiply\epsftmp\epsftsize \advance\epsfrsize-\epsftmp
     \epsftmp=\epsfxsize
     \loop \advance\epsfrsize\epsfrsize \divide\epsftmp 2
     \ifnum\epsftmp>0
        \ifnum\epsfrsize<\epsftsize\else
           \advance\epsfrsize-\epsftsize \advance\epsfysize\epsftmp \fi
     \repeat
     \epsfrsize=0pt
    \else
     \epsfrsize=\epsfysize
    \fi
   \fi
   \ifepsfverbose\message{#1: width=\the\epsfxsize, height=\the\epsfysize}\fi
   \epsftmp=10\epsfxsize \divide\epsftmp\pspoints
   \vbox to\epsfysize{\vfil\hbox to\epsfxsize{%
      \ifnum\epsfrsize=0\relax
        \includegraphics{#1}%
      \else
        \epsfrsize=10\epsfysize \divide\epsfrsize\pspoints
        \includegraphics{#1}%
      \fi
      \hfil}}%
\global\epsfxsize=0pt\global\epsfysize=0pt}%
{\catcode`\%=12 \global\let\epsfpercent=
\long\def\epsfaux#1#2:#3\\{\ifx#1\epsfpercent
   \def\testit{#2}\ifx\testit\epsfbblit
      \epsfgrab #3 . . . \\%
      \epsffileokfalse
      \global\epsfbbfoundtrue
   \fi\else\ifx#1\par\else\epsffileokfalse\fi\fi}%
\def\epsfempty{}%
\def\epsfgrab #1 #2 #3 #4 #5\\{%
\global\def\epsfllx{#1}\ifx\epsfllx\epsfempty
      \epsfgrab #2 #3 #4 #5 .\\\else
   \global\def\epsflly{#2}%
   \global\def\epsfurx{#3}\global\def\epsfury{#4}\fi}%
\def\epsfsize#1#2{\epsfxsize}
\begin{document}

\title{SU(3) vortex-like configurations in the maximal center gauge
\vskip-3cm\hfill\small FTUAM-99-24\vskip2.6cm}

\author{A. Montero\address{ Departamento de F\'{\i}sica Te\'orica C-XI,
        Universidad Aut\'onoma de Madrid, Madrid 28049, Spain} }

\begin{abstract}
A new algorithm for fixing the gauge to (direct) maximal center
gauge in SU(N) lattice gauge theory is presented. We check how 
this method works on SU(3) configurations which are vortex-like,
and show how these configurations look like when center projected.

\end{abstract}

\maketitle

\section{Introduction.}
\vspace{-0.25 cm}
The maximal center gauge is widely used to study low energy
phenomena such as the confinement property or the breaking of the
chiral symmetry, as can be seen in a set of recent works 
\cite{green1,green2,green3,lang,lang2,cher1,cher2,forcrand,steph} and
also in these proceedings. Nevertheless, at least to our knowlegde,
there was no efficient method of direct gauge fixing to maximal center 
gauge in SU(N) lattice gauge theory for values $N>2$. This is the motivation
of the first part of this work in which we present a new algorithm of
direct center gauge fixing to maximal center gauge. We apply this new method
to previously prepared SU(3) vortex-like configurations.
Our purpose in the second part of this work is 
to know how these configurations look like in the maximal center gauge, and therefore, 
whether these solutions have the properties described in references 
\cite{green1,green2,green3,lang,lang2,cher1,cher2,forcrand,steph} for a confining
object. For a more detailed description of our results see reference \cite{mont1}.
\vspace{-0.30 cm}
\section{The method.}
\vspace{-0.25 cm}
   The maximal center gauge in SU(N) lattice gauge theory is defined as the
gauge which brings link variables $U$ as close as possible to elements of the center
$Z_N = \left\{e^{2\pi m i/N}\Id \right.$ , $\left. m=0,...,N-1 \right\}$.
This can be achieved by maximizing the following quantity:
\vspace{-0.2 cm}
\bea
R=\frac{1}{N_{site}N_{dim}N^2}\sum_{n,\mu}\mid\mbox{Tr}\;U_\mu(n)\mid^2  
\eea
\vspace{-0.4 cm}

\noindent
where $N_{site}$ is the number of sites in the lattice and $N_{dim}$ the number of 
dimensions ($R$ satisfies $|R| \! \le \! 1$). Our procedure is based on a local update
of $R$. We maximize this quantity respect to a gauge transformation $G(n)$ defined
at the lattice point $n$. We build the SU(N) matrix $G(n)$ from a SU(2) matrix $g(n)$
as in the Cabbibo-Marinari-Okawa method. Doing the update in this way 
the problem is reduced to finding the maximum of a quadratic form, and this can be done using
standard methods (see \cite{mont1} for details).  Once we obtain the SU(2) matrix $g(n)$ 
maximizing $R$, we update the $U_{\mu}$ variables touching the
site $n$: $U_{\mu}(n) \rightarrow G(n)U_{\mu}(n)$ and
$U_{\mu}(n-\hat{\mu}) \rightarrow U_{\mu}(n-\hat{\mu})G^\dg(n)$.
We repeat this procedure over the $N(N-1)/2$ SU(2) subgroups of SU(N) and
over all lattice points. When the whole lattice is covered once we say
we have performed one center gauge fixing sweep. We make a number of
center gauge fixing sweeps on a lattice configuration and stop the
procedure when the quantity $R$ is stable within a given precision
($10^{-8}$ in this work).

\vspace{-0.30 cm}
\section{The vortex solution.}
\vspace{-0.25 cm}
Following the same methods used in \cite{mont2} for the SU(2) group,
we built a SU(3) Yang-Mills configuration which has the following features:
\vspace{-0.2 cm}
\begin{itemize}
\item It is a solution of the SU(3) Yang-Mills classical equations in $R^4$.
\vspace{-0.25 cm}
\item It is constructed from $R^2 \times T^2$ solutions of the SU(3)
Yang-Mills classical equations by glueing to themselves the two periodic directions 
($R^2 \times T^2$ is considered the limit of $T^4$ with two periods much larger than the
other two).
\vspace{-0.25 cm}
\item The $R^2 \times T^2$ building block satisfies non-orthogonal twisted
boundary conditions, is (anti)self-dual and has minimal action $S=8\pi^2/3$.
The crucial point for this building block to have vortex-like properties is that 
the twist in the small plane must be non trivial.
\end{itemize}

  We isolate the $R^2 \times T^2$ building block using the lattice aproach. 
Working on lattices of sizes $N_l \! \times \! N_l \! \times \! N_s \! \times \! N_s$
(directions $t,x,y,z$ respectively and $N_l \! >> \! N_s$) and imposing twisted
boundary conditions given by the twist vectors $\vec{k} \! = \! \vec{m} \! = \! (1,0,0)$ we 
isolate SU(3) configurations with topological charge $Q \! = \!-1/3$ and minimal action
$S=8\pi^2/3$ using a standard cooling algorithm.
The lattice sizes used are $N_s=4,5,6$ and $N_l=6 \times N_s$, and we 
obtain configurations with action $S/(8\pi^2/3) = 0.92219, 0.95015$, $0.96536$
respectively. We fix the length in the small directions equal to $1$ and
then the lattice spacing is $a=1/N_s$.

\vspace{-0.25 cm}
\section{Properties of the solution}
\vspace{-0.25 cm}
{\bf A)} From the field strength ${\bf F}_{\mu \nu}$, obtained using the clover
average of $1 \times 1$ plaquettes, we calculate the quantity
$\Epsilon_{\mu \nu}(x_{\mu},x_{\nu})$, the integral of the action density in two
directions, defined as,
\vspace{-0.15 cm}
\bea
\Epsilon_{\mu \nu}(x_{\mu},x_{\nu}) = \int d^2 x_{\alpha \neq \mu, \nu} \hspace{0.2 cm} \frac{1}{2} 
\Tr \left( {\bf F}_{\rho \sigma} {\bf F}_{\rho \sigma} \right) 
\eea

\vspace{-0.15 cm}
\noindent
and obtain the following results:
\begin{itemize}
\vspace{-0.20 cm}
\item Localized in the $xt$ plane as shown in the plot of $\Epsilon_{xt}(x,t)$ included
in figure 1.
\vspace{-0.25 cm}
\item Localized in the $x$ direction and almost flat in the
$y$ direction as shown in the plot of $\Epsilon_{xy}(x,y)$ included in 
figure 1. In this case we repeat six times the y direction to obtain a 
box size equal to the one used in the figure for $\Epsilon_{xt}(x,t)$. Identical
figures are obtained changing $x \rightarrow t$ and $y \rightarrow z$.
\vspace{-0.25 cm}
\item Almost flat in the $y,z$ directions as shown in the 
plot of $\Epsilon_{yz}(y,z)$ included in figure 1. We repeat six times both
directions to obtain a box size equal to the one used for $\Epsilon_{xt}(x,t)$. 
\end{itemize}
\vspace{-0.2 cm}
\begin{figure}
\vspace{5.0cm}
\includegraphics{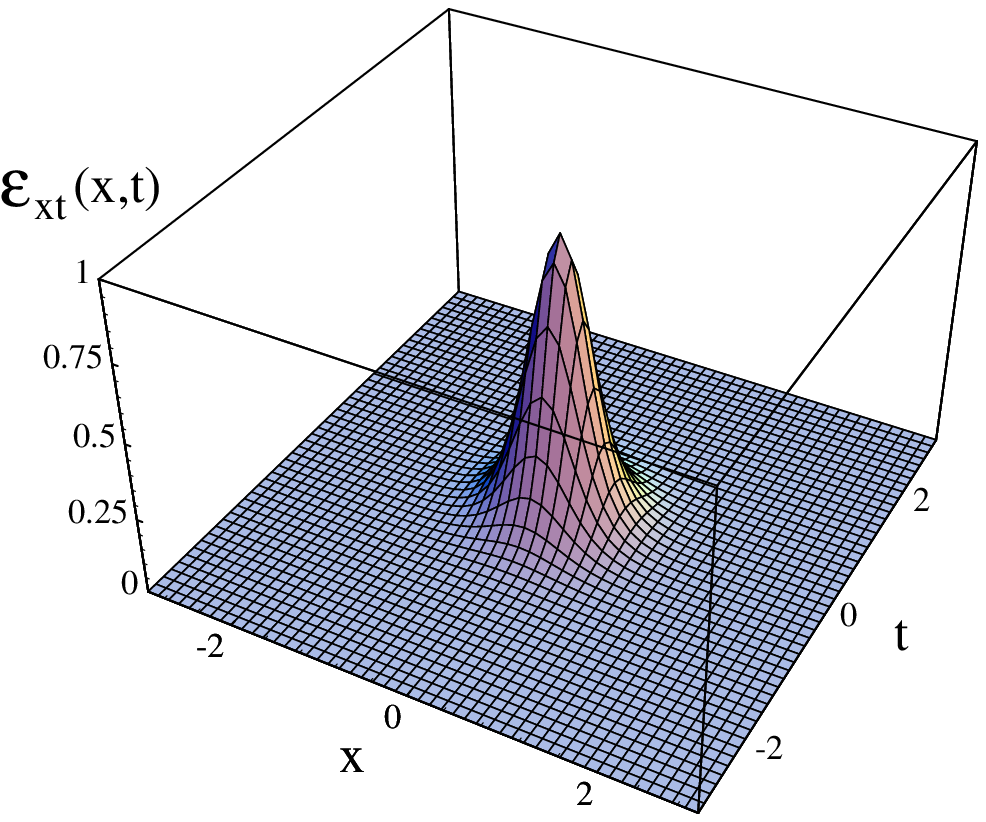}
\includegraphics{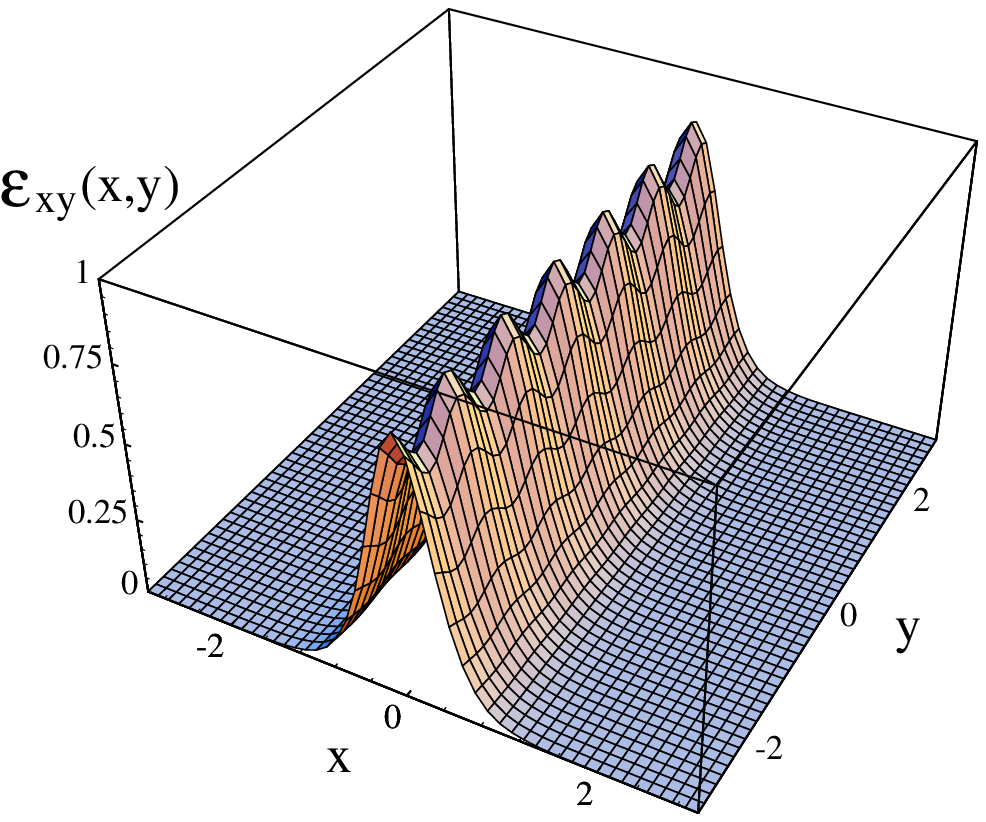}
\includegraphics{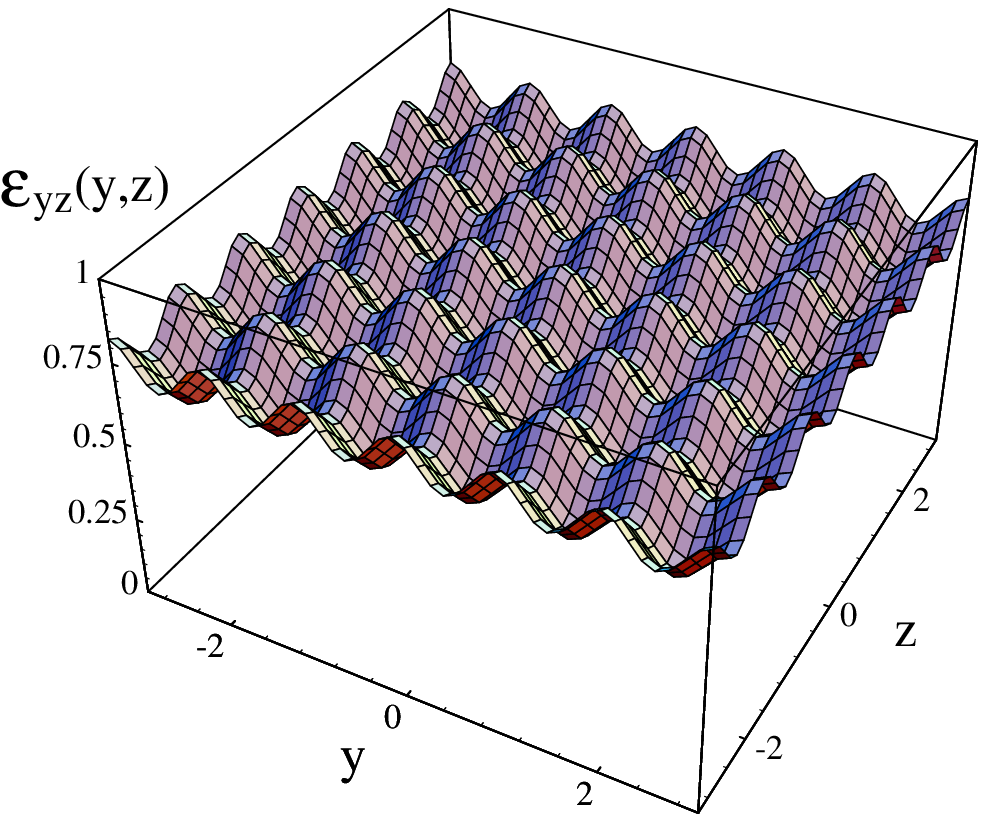}
\vspace{2.5cm}
\caption{Action densities integrated in two directions.}
\vspace{-0.5 cm}
\end{figure}
{\bf B)} We calculate the Wilson loop $W_C(r)$ around this object,
\vspace{-0.15 cm}
\bea
W_C(r) = \frac{1}{N} \hspace{0.1 cm} \Tr \hspace{0.1 cm} 
         \left( Pexp \int_C \imath A_{\mu} dx_{\mu} \right) 
         \label{eq:wl}
\eea

\vspace{-0.2 cm}
\noindent
being $C$ an $r \times r$ square loop in the $xt$ plane centered at the maximum of the solution. 
We parametrize $W_C(r)$ by the functions $L(r)$ (its module) and $\phi(r)$ (its phase). 
In figure 2 we show the quantities $L(r)$ and $\phi(r)$ as a function of $r$, putting in the
same plots the values with the square loop centered at the minimum of 
the solutions in the $y,z$ directions and at the maximum in the $t,x$ directions.
The conclusions extracted from figure 2 are the following:
\begin{itemize}
\vspace{-0.15 cm}
\item These functions are almost independent of coordinates $y$ and $z$.
\vspace{-0.25 cm}
\item When $r$ is bigger than the size of the object we obtain
$W_C(r) \sim exp(i 4\pi /3) $, an element of the group center.
\end{itemize}
\vspace{-0.25 cm}
These are the two expected properties for a vortex.

\begin{figure}[htb]
\vspace{7.3 cm}
\includegraphics{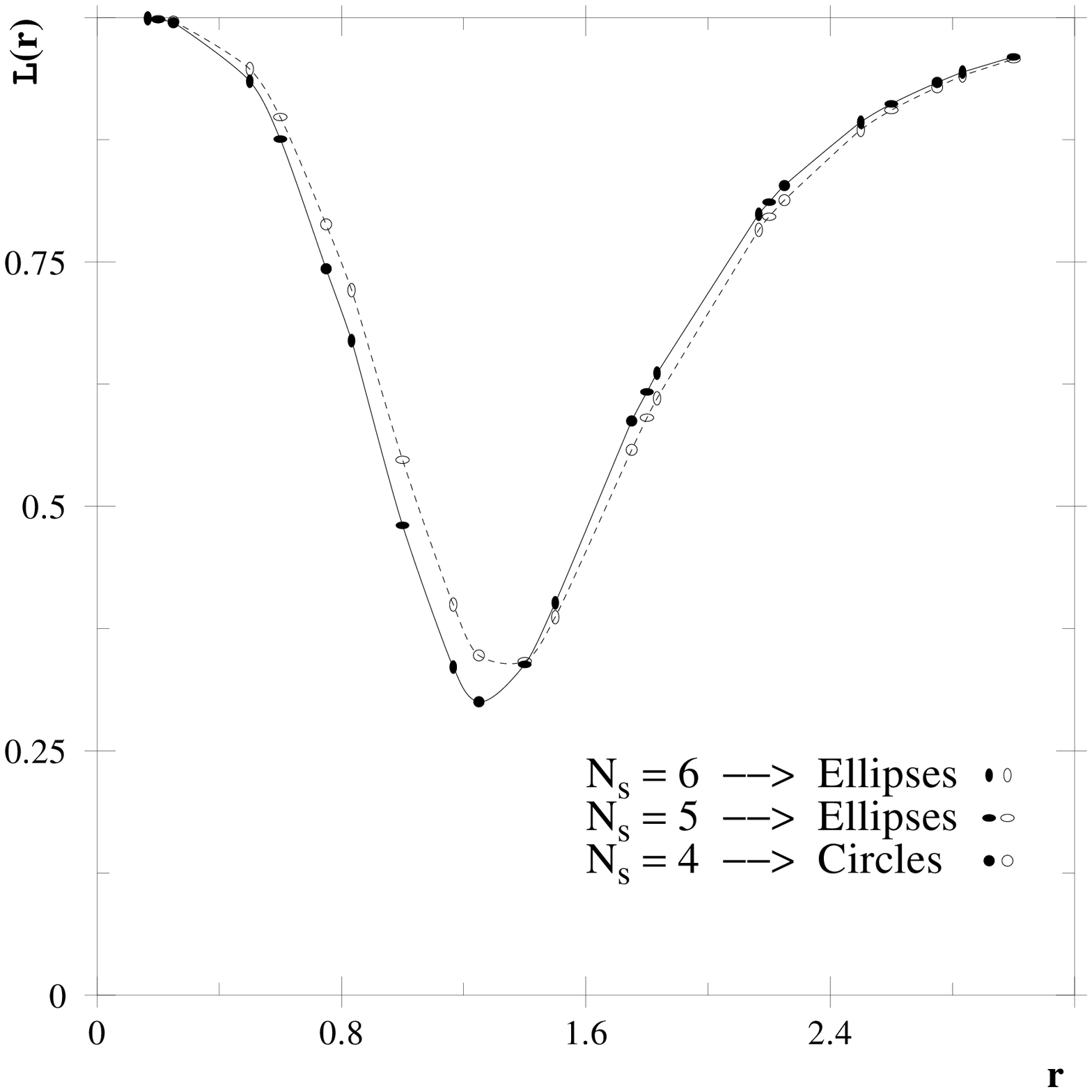}
\includegraphics{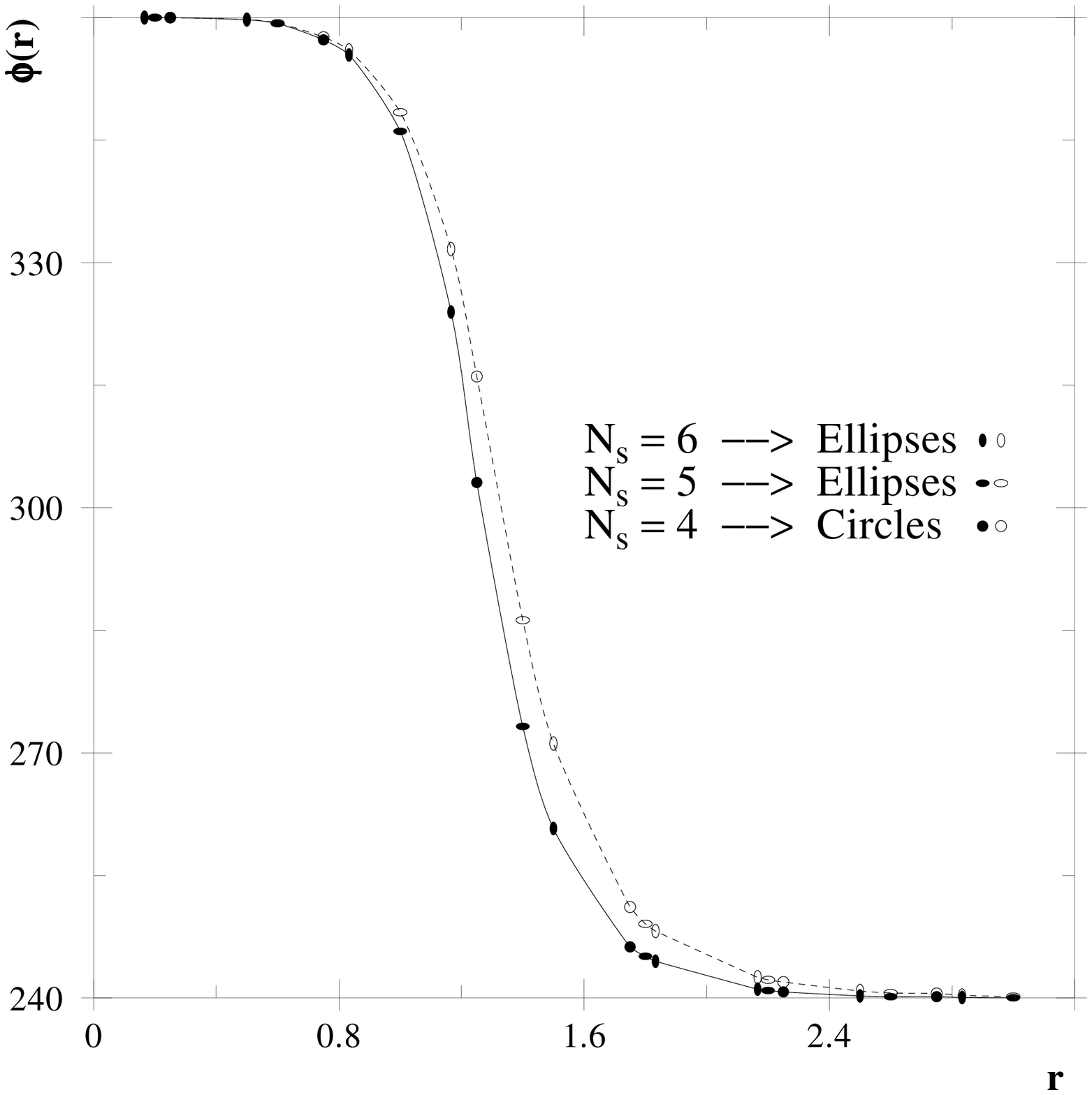}
\vspace{3.2 cm}
\caption{The module $L(r)$ (top figure) and the phase $\phi(r)$
(bottom figure) of the Wilson loop defined in equation \ref{eq:wl} as a function of $r$.
Full symbols correspond to Wilson loops centered at the maximum of the solutions 
and empty symbols to the same quantity centered at the minimum of the solutions
in the $y,z$ directions and at the maximum in the $t,x$ directions.}
\vspace{-0.85 cm}
\end{figure}

\vspace{-0.25 cm}
\section{The center-projected configuration.}
\vspace{-0.25 cm}
We apply to the obtained solutions the method of center gauge fixing. 
For the SU(3) vortex-like solutions we have worked with we have obtained 
the following maximum values: $R=0.90123970$, $0.93246431$ and $R=0.94918155$,
for the lattices sizes: $N_s=4,5$ and $N_s=6$ respectively.
We project the SU(N) link variables to $Z_N$ link variables 
and we calculate the values of the plaquettes from the $Z_N$ link
variables. We obtain the following results:
\begin{itemize}
\vspace{-0.20 cm}
\item In the $xt$ plane, the plane in which the vortex is localized, we 
obtain  for the configurations with maximum value of $R$ the same structure
for all $y,z$ points. Only two plaquettes are different from the identity.
The first one is located at the top-right corner and reflects the 
use of twisted boundary conditions. 
The other one is located near the maximum in the action density 
of the solution and has the same value of the Wilson loop shown before when the 
size of the loop is much bigger than the size of the solution. This is one of the 
most interesting results of our work, showing how the vortex properties are 
reflected in the center-projected configuration.   
\vspace{-0.30 cm}
\item In the $xy$ plane ( or $xz,ty,tz$ plane). In this plane the vortex is localized in one 
direction and almost flat in the other. There is no regular structure for all $t,z$ points. The
most common configuration is that one with all plaquettes equal to the identity.
\vspace{-0.25 cm}
\item In the $yz$ plane. The solution is almost flat in both directions. As in the previous 
case, there is no regular structure for all $t,x$ points. The
most common configuration is that one with a
very similar  structure to the one in the $xt$ plane.
\end{itemize}
\vspace{-0.60 cm}
\section{Conclusions.}
\vspace{-0.25 cm}
The main conclusions of this work are the following two. First,
our method of gauge fixing to maximal center gauge works very well.
And second, we have seen that, for the configurations with maximum
value of $R$, there is a clear relation between the structure of the
SU(3) configuration and the structure of the center-projected one.

{\bf Acknowledgments}:  The present work was financed by CICYT 
under grant AEN97-1678. I acknowledge useful conversations with
M. Garc\'{\i}a P\'erez, A. Gonz\'alez-Arroyo  and C. Pena.

\end{document}